# Spatial parking planning design with mixed conventional and autonomous vehicles

Qida SU, David Z.W. WANG[*]

School of Civil and Environmental Engineering, Nanyang Technological University, 50 Nanyang Avenue, Singapore 639798

**Abstract**

Travellers in autonomous vehicles (AVs) need not to walk to the destination any more after parking like those in conventional human-driven vehicles (HVs). Instead, they can drop off directly at the destination and AVs can cruise for parking autonomously. It is a revolutionary change that such parking autonomy of AVs may increase the potential parking span substantially and affect the spatial parking equilibrium. Given this, from urban planners' perspective, it is of great necessity to reconsider the planning of parking supply along the city. To this end, this paper is the first to examine the spatial parking equilibrium considering the mix of AVs and HVs with parking cruising effect. It is found that the equilibrium solution of travellers' parking location choices can be biased due to the ignorance of cruising effects. On top of that, the optimal parking span of AVs at given parking supply should be no less than that at equilibrium. Besides, the optimal parking planning to minimize the total parking cost is also explored in a bi-level parking planning design problem (PPDP). While the optimal differentiated pricing allows the system to achieve optimal parking distribution, this study suggests that it is beneficial to encourage AVs to cruise further to park by reserving less than enough parking areas for AVs.

---

[*] Corresponding author. Tel: (65) 6790-5281

Email: wangzhiwei@ntu.edu.sg   qida.su@ntu.edu.sg

1. **Introduction**

Autonomous vehicles (AVs) are widely recognized as the future of urban mobility (Schreurs & Steuwer, 2015). With autopilot systems, individual drivers no longer need to pilot themselves, allowing them to spend their time *en route* on other activities in a fully automated vehicle. In particular, unlike the Human-driven vehicles (HVs), drivers can drop off directly after arriving at the destination and leave the AVs to self-cruise for parking.

In fact, on the way to vehicle automation, the Society of Automotive Engineers (SAE) has defined six levels of vehicle autonomy, from Level 0 with no driving automation to Level 5 with full driving automation ("Taxonomy and Definitions for Terms Related to Driving Automation Systems for On-Road Motor Vehicles," 2018). Indeed, there will be no doubt that full autonomy technologies for vehicles will be achieved in the coming age (Luettel et al., 2012). In view of the promising prospect of autonomous vehicles as urban mobility solution, it is imperative to better understand the emerging travel behaviours of AV travellers, such as the autonomous cruising for parking, so as to propose more efficient traffic management measures in the presence of prevalent AVs.

On this account, it is of pressing importance to prepare for the coming era of AVs in advance with up-to-date management measures (Fagnant & Kockelman, 2015). In fact, research attentions have been paid to the management of AVs in recent literature regarding intersection controls (Naumann et al., 1998; Perronnet et al., 2013; Levin et al., 2017; Yu et al., 2019), safety issues (Fernandes & Nunes, 2012; Kalra & Paddock, 2016; Shladover & Nowakowski, 2019), road tolls (Sharon et al., 2017;

Simoni et al., 2019; Tscharaktschiew & Evangelinos, 2019) , supply of AVs (van den Berg & Verhoef, 2016; Chen et al., 2020), reservation of AVs (Lamotte et al., 2017), the shared-AVs (SAVs) (Fagnant & Kockelman, 2018; Tian et al., 2019), dedicated AV lanes (Chen et al., 2016; Chen et al., 2017; Ghiasi et al., 2017; Movaghar et al., 2020), car-parks design (Nourinejad et al., 2018), etc.

More recently, some research works have noticed the autonomous parking behaviour of AVs. Liu (2018) first investigated the parking with AVs in the morning commute problem with time-varying congestion, and later it was extended to the day-long context (Zhang et al., 2019b). Su and Wang (2020) explored the parking location choices of AV commuters from different residential clusters considering the distant parking options. However, 100% market penetration of AVs is assumed in most of the previous work without the consideration of the conventional HVs. While the full penetration of AVs requires a long transitional period, studying the mixed case with coexistence of AVs and HVs would be a more timely issue for policymakers. Besides, the negative externalities of AVs in parking cruising were scarcely taken into account except in (Zhang et al., 2019a; Levin et al., 2020). Due to parking autonomy, parking demand can shift from downtown to adjacent neighbourhoods (Zhang & Guhathakurta, 2017). That is to say, the AV's parking span could be much larger than that of HVs. Such a larger parking span of AV boosts the total distance on parking cruising, which can, in turn, exacerbate congestion on road traffic (Fagnant et al., 2015).

While many cities in the world are envisioning future mobility system with mixed AVs and HVs, the urban planners need to understand how to best plan the parking spaces citywide with explicit consideration of the different parking behaviours of AVs and HVs. Granted that the land spaces in city centre are limited with exorbitant prices, one fundamental question is that, whether more parking spaces should be shifted outwards from the city centre considering longer parking span and smaller unit size of parking space in AVs. Besides, in the parking facility at one specific location, what is the optimal parking space allocation for AVs and HVs? If differentiated pricing at different parking locations for different vehicle types, i.e., HV and AV, is imposed, what is the optimal pricing scheme? To answer these questions is practically necessary for urban planners to prepare optimal parking planning citywide for the introduction of AVs into future urban transportation system.

Indeed, there are plenty of studies on the transportation planning incorporating the strategic, tactical, and operational decisions of regulators, which is called the urban transportation network design problem (UTNDP) (Boyce & Janson, 1980; Magnanti & Wong, 1984; Friesz, 1985; Yang & H. Bell, 1998; Guihaire & Hao, 2008; Wang & Lo, 2010; Farahani et al., 2013; Szeto et al., 2015). Research interest in UTNDP has mainly been paid to road network design (RNDP) and public transit network design (PTNDP). RNDP usually concerns about building new roads, expanding the existing roads or determining lane allocation, whereas PTNDP often emphasizes on the service frequency and the transit schedule. Nevertheless, in the fragmentary studies in UTNDP related to parking, they focus more on the park-and-ride facilities, rather than the parking supply along the city (Lam et al., 2001; García & Marín, 2002; Du & Wang, 2014). To the best of our knowledge, previous studies have seldom examined the citywide design issue on parking planning, due to the confined span of parking location choice for HVs. Only in (Levin et al., 2020), the network parking infrastructure design was examined in an extension of (Zhang et al., 2019a), incorporating the impact of the different AV market penetration.

Admittedly, the AV future urges a citywide parking planning design. This paper then aims to fill this research gap by addressing the aforementioned questions on optimal parking supply when planning for the future transportation system with mixed traffic. While (Levin et al., 2020) applied metaheuristic solution method for the network parking design problem focusing on the additional parking spaces for AV repositioning, we target to first theoretically scrutinize the parking choice difference of AV and HV as well as the resulting parking land usage difference along the city. We then intend to provide indicative insights on spatial parking planning to analytically determine the optimal parking intensity and allocation for both AVs and HVs. To achieve so, we investigate the parking location choice behaviours of mixed traffic travellers in a linear travel corridor while parking cruising effects are clearly captured. A continuum modelling approach is applied to model and solve the spatial equilibrium of travellers' parking location choices. Based on this, policy indications in parking planning are provided.

To be specific, we hereinafter refer to those private-owned vehicles in the driving automation of Level 4&5 with the capability of self-driving as AVs, and refer to those with human driving during parking cruising as HVs.

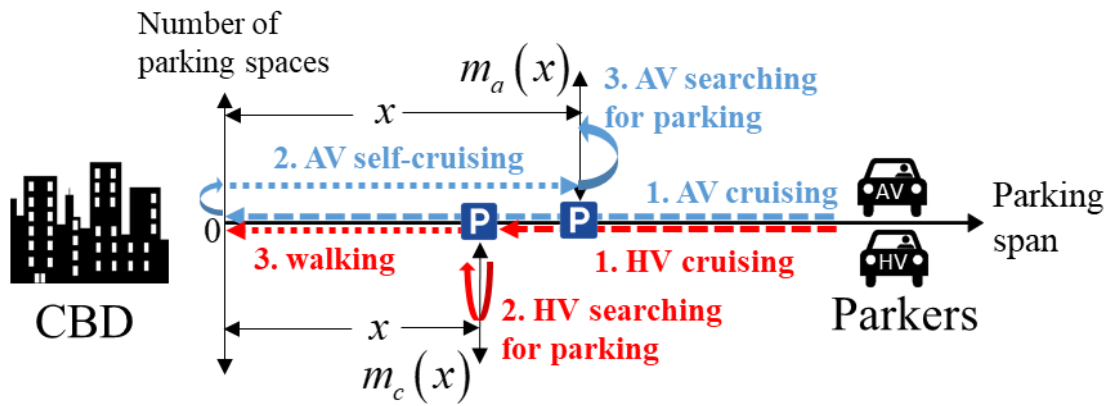

Figure 1 The parking processes of AVs and HVs in a linear city corridor.

First and foremost, the question of how travellers in AVs and HVs behave differently in their parking locations' choices is thoroughly addressed considering cruising effects. As was done in Anderson and de Palma (2004), a linear travel corridor with a paralleled two-way arterial road is considered (Figure 1), with $x$ denoting the distance to the CBD (located at $x=0$). Parking spaces are distributed along the travel corridor with total parking areas $k(x)>0$ at location $x$. It should be noted that $k(x)$ may vary non-linearly with $x$, depending on the parking supply in practice. In addition, it is assumed that the parking spaces for AVs and HVs are separated and independent of each other,[1] such that the searching for parking space of AVs and HVs are independent at any location $x$, and the externalities imposed upon the other type of vehicles exist on the arterial road only. Further, the unit size of parking space is assumed to be 1 for HVs and $\phi \in (0,1]$ for AVs, which is constant along the city. Let $\theta(x) \in [0,1]$ denote the proportion of parking areas allocated to AVs at $x$, and the number of parking spaces becomes $m_a(x) = \theta(x)k(x)/\phi$ and $m_c(x) = (1-\theta(x))k(x)$. It is also assumed that there are on average $N$ commuters in the peak hour who live at the suburban area. Each of them drives to CBD to work and parks his/her car without parking reservation. Among them $\varepsilon \times 100\%$ are traveling with AVs $(N_a = \varepsilon N)$ and the rest are in HVs $(N_c = (1-\varepsilon)N)$. Travellers in HVs cruise towards city center to determine their parking location. Parking at locations closer to CBD requires less walking distance, however, at the cost of longer cruising distance and searching time for a vacant parking space. Contrarily, travellers in AVs drop off at city center first and let the vehicles self-cruise outward from city center to search for parking space. The further they park, the longer distance they need to cruise with larger energy consumption (electricity/gasoline), for the sake of easier parking. We notice that previous research on AVs scarcely took parking cruising into account, which may lead to wrong estimation on AV's parking span in urban area, and hence oversupplied or insufficient parking space provision, particularly in the environment of mixed AVs and HVs traffic. Thus, this study explicitly considers the cruising effects in analyzing travellers' parking location choice behaviour with mixed traffic.

---

[1] This assumption is made based on the prospect of different parking spaces dimensions/layouts and different parking behaviours of AVs and HVs (Nourinejad et al., 2018). Compared to HVs, AVs require less space in a single parking slot, can search parking in platooning, and are usually electric vehicles (EVs) that need charging facilities during parking. Thus, it is reasonable to assume such parking independence. One analogous observation in practice is that current garages usually allocate a dedicated parking and charging area for electric vehicles (EVs). In fact, the parking independence of AVs and HVs also allows policymakers to achieve more sophisticated managerial goals, e.g., to force AVs to park further and leave more spaces for HVs so as to reduce walking, or to save the parking land use with more dedicated AV parking spaces (which are in smaller size).

In order to minimize the total parking cost of travellers, a bi-level parking planning design problem (PPDP) is next developed to scrutinize the optimal parking pricing and urban parking planning design. Here, the parking planning design is mainly referred to as the land scale $k(x)$ and the allocation among different vehicle types $\theta(x)$ along the city. At the lower level problem (LLP) with given parking design, the parking span of AVs is found to be tighter at optimum due to the internalization of cruising and searching for parking, in line with the results in literature. In addition, the optimal differentiated parking pricing is determined depending on the vehicle type and the parking location. Nevertheless, while optimal differentiated pricing optimizes the respective spatial distribution of parking choices, we notice that the total parking cost can be further reduced at the upper level problem (ULP) with appropriate parking planning. With limited expenditure on parking infrastructure, the optimal parking planning design for every given AV penetration is determined under mild assumptions. Regulators can then re-evaluate or design in advance the guidelines for existing or future parking facilities on the parking land scale and allocation. In general, it is socially beneficial to encourage AVs to cruise further to park by allocating less than enough parking spaces to AVs.

The rest of the paper is organized as follows. First, Section 2 continues to develop the model framework with mixed AVs and HVs in a linear parking corridor model. Section 3 then studies the unpriced spatial equilibrium of parking location choices. In the bi-level PPDP, the optimal parking choice distribution at LLP is first examined in Section 4, followed by the exploration of optimal parking planning at ULP in Section 5. Numerical examples are next presented in Section 6. Finally, Section 7 wraps up the paper with some summaries and future research directions.

2. **Model formulation**

We firstly continue to work on the parking model formulation and the list of notation throughout the paper is presented in Appendix A. As introduced above, the key factors on the parking spaces' supply to vehicle type $i(=a,c)$ at location $x$ are $k(x)$ and $\theta(x)$. In general, $k(x)$ and $\theta(x)$ are determined by the urban planners to design how many parking spaces should be built and how many of them should be allocated for AV parking at that location. More intuitively, one may consider a garage located at any location along the corridor continuously with total parking area $k(x)$, where $\theta(x)$ of them are dedicated for AVs only and the other are for HVs.

As for the different parking behaviours of AVs and HVs, as shown in Figure 1, the parking process of HVs is the same as that in Anderson et al. (2004): 1. Travel towards the city centre and determine the location to park $\rightarrow$ 2.Find a parking space at that location $\rightarrow$ 3.Walk to the CBD. Comparatively, the parking process for travellers in AVs would be, however, different from those in HVs due to the self-driving capability, which follows: 1. Travel towards CBD directly and drop off $\rightarrow$ 2. Self-travel outwards from the city centre and determine the location to park $\rightarrow$ 3. Find a parking space at that location.

In fact, the parking location choices for travellers are determined by not only the parking process but also the leaving process from parking. After shopping/doing business at CBD, the process of HVs to leave follows the procedure: 1. Walk from CBD to parking location $\rightarrow$ 2. Travel outwards to home. And the process of AVs to leave CBD follows the procedure: 1. Self-travel inwards to CBD $\to$ 2. Pick up the parker and travel outwards. It can be readily found that the process to leave in reverse is exactly the process to park for both types of vehicles, except the searching at the selected parking location. Therefore, though the parking location is a composite decision making of both parking and leaving, we hereinafter only consider the parking process only. Before we formulate the generalized parking costs of both types of vehicles in Section 2.4, several parking behaviours are modelled in the following.

## 2.1 HV's walking and AV's self-driving

Conventionally, a traveller in HV parking at location $x$ needs to walk for a distance $(= x - 0)$ towards the CBD, with walking cost $\lambda_c x$ imposed, where $\lambda_c$ is the walking cost per unit distance. The walking from the parking lot to the arterial road is ignored for simplicity. However, travellers in AVs no longer need to walk. Instead, AV drives itself outwards for a distance $x$ to its parking location with self-driving cost $\lambda_a x$, where $\lambda_a$ is the aggregate cost per unit distance of AV's self-driving on the arterial road, which may combine the fuel/electric cost and other depreciation costs. The cost of drop-off for AV passengers is assumed to be zero. Naturally, walking is more unfavourable and hence we let $\lambda_c \gg \lambda_a$. Also, we use $\overline{x_i}$ to denote the parking spans (the furthest parking locations from the CBD) of vehicle type $i (= a, c)$. In other words, $\overline{x_i}$ is the longest distance a HV traveller is willing to walk, or the longest distance an AV is willing to self-drive. With $\overline{x_i}$, the maximum costs of walking for HVs and self-driving for AVs on the arterial road are $\lambda_c \overline{x_c}$ and $\lambda_a \overline{x_a}$, respectively.

## 2.2 Parking spot searching

Parking spot searching is referred to as the searching for a parking spot after the traveller determines the parking location and exits from the arterial road. Once a vehicle arrives at its selected parking location, it needs to search for a vacant spot within the allocated parking spaces of its type (AV or HV). For both types of vehicles, we assume that it takes more time to find a vacant spot at any location when there is higher parking occupancy of the same vehicle type, which is the actual parking spaces occupied over the allocated parking spaces $m_i(x)$ $i(=a,c)$. While vehicles in the location with low parking occupancy can find a parking space easily with shorter searching time, those in the location with high parking occupancy can bear the dramatically increasing searching time (Axhausen et al., 1994; Horni et al., 2013; Levy et al., 2013; Qian & Rajagopal, 2014; Inci & Lindsey, 2015). Thus, the expected searching time for parking of an individual in vehicle type $i(=a,c)$ at location $x(\geq 0)$, $S_i(x)$, should satisfy

$$\frac{\partial S_i(x)}{\partial n_i(x)} > 0, \quad \frac{\partial^2 S_i(x)}{\partial n_i^2(x)} \geq 0, \quad (i=a,c), \tag{1}$$

where $n_i(x) \in [0, m_i(x)]$ denotes the distribution of parking location choices (or parking distribution).

Such first- and second-order derivatives with respect to $n_i(x)$ in Eq.(1) ensure the monotonicity and convexity of $S_i(x)$. Obviously, when $n_i(x)=0$, $S_i(x)$ is minimized. For simplification purpose, it is assumed that such minimum parking searching time is equal for both type of vehicles, i.e., $S_i(x)\big|_{n_i(x)=0} = S_{\min}$. Letting $\gamma_i$ denote the unit time cost of parking searching, we can represent the searching cost in type $i$ at location $x$ as $\gamma_i S_i(x)$ $(i=a,c)$. Hereby, as we are working on a long-term planning problem, the searching time cost is indeed an average cost for search a parking spot in the long run, rather than that for one specific day with temporal factors incorporated. It is worth noting that while the drivers in HVs need to steer the vehicle to the parking spaces even with full parking information, AVs can finish the process by itself without driver and hence we let $0 < \gamma_a \ll \gamma_c$. Besides, the conservation of parking demand requires $n_i(x)$ to satisfy that

$$\int_0^{\bar{x}_c} n_c(x)dx = N_c \text{ and } \int_0^{\bar{x}_a} n_a(x)dx = N_a. \tag{2}$$

*2.3    Parking cruising on the arterial road*

Though the searching for parking spots after determining the parking location for the two types of vehicles is assumed to be independent of each other, they share the arterial road usage during the cruising for parking. Basically, those vehicles when cruising for parking would slow down and make a turn to exit the arterial road, imposing negative congestion externalities to the other cruising vehicles. To formulate such cruising effect on the arterial road, we assume that the induced delay for one individual cruising vehicle at location $x$ is a linearly increasing function of the number of vehicles cruising for parking in the same direction in the small interval $[x, x+\Delta x]$ (Anderson et al., 2004), i.e., $\sum_{i \to \text{same direction}} \beta_i n_i(x)\Delta x$. Here, $\beta_i$ is the given coefficient for each type $i(=a,c)$ with $\beta_a < \beta_c$. Additionally, we assume $\beta_c < \lambda_c / k_{\max}$ to ensure the outward parking preference of HVs.[2] In particular, when $\beta_i = 0$, it reduces to the case without considering the cruising effects as in previous literature (Anderson et al., 2004), which will be used as a benchmark case for comparison in next sections.

---

[2] In fact, when the cruising effect dominates in travel cost, due to the inward cruising direction of HV travellers, they prefer to park far away and walk to CBD to avoid congestion, which however, leads to large total walking cost to the system. As shown later in Lemma 3.2, this assumption ensure $n_c^{e\,\prime}(x) < 0$ and all HV travellers tend to park close to CBD.

Therefore, for HVs, they only cruise inwards to CBD, while the outward cruising of AVs does not affect the inward HVs. The average cruising cost for one HV traveller who chooses to park at $x$ would be:

$$c_c(x) = \beta_c \int_x^{\bar{x}_c} n_c(u) du. \tag{3}$$

On the contrary, AVs would first travel directly to CBD along with the cruising HVs, and then self-cruise outwards to the parking location. The average cruising cost for an AV traveller who chooses to park at $x$ is

$$\begin{aligned} c_a(x) &= \beta_c \int_0^{\bar{x}_c} n_c(u) du + \beta_a \int_0^x n_a(u) du \\ &= \beta_c N_c + \beta_a \int_0^x n_a(u) du. \end{aligned} \tag{4}$$

Here, the first term in Eq.(4) represents the cruising delay resulted from cruising HVs in the inward direction and the second term delineates the cruising delay resulted from AVs when cruising outward from CBD to the selected parking location.

*2.4  Generalized travel cost*

Based on the analysis above, the generalized parking cost for travellers in HVs and AVs choosing to park at location $x$, including walking/self-driving cost, parking spot searching cost, and cruising delay cost on the arterial road, can be simply defined as follows:

$$P_i(x) = \lambda_i x + \gamma_i S_i(x) + c_i(x), \quad (i = a, c). \tag{5}$$

So far, the pricing of parking has not been incorporated in the generalized cost formulation (or a uniform pricing is assumed at each location). This assumption would be relaxed in Section 4 and 5 with the aim to optimize travellers' spatial parking distribution. In the following, we will first examine the unpriced spatial equilibrium.

3. **Unpriced spatial equilibrium of parking location choices**

Consider a linear travel corridor with mixed HVs and AVs. The spatial equilibrium of parking location choices in the corridor without considering the differentiate parking price would be derived in this section by applying continuum modelling approach.

Evidently, at equilibrium, all travellers in the same type of vehicles share the equal generalized parking cost at any used parking location and no one can reduce his/her parking cost by changing the parking location unilaterally, i.e.,

$$P_i'(x) = 0, P_i(x) = p_i.  \quad (6)$$

Later in Section 3.1, the derivation of parking equilibrium is demonstrated with specifying the parking searching time functions. Nevertheless, without specifying $S_i(x)$, some general properties at spatial equilibrium can still be observed, as stated in some lemmas.

Foremost, the trend of equilibrium parking distribution $n_i^e(x)$ (the superscript $e$ indicates the unpriced spatial equilibrium) is found to be similar for both vehicle types.

**Lemma 3.1** *When $N_i > 0$, each parking location along the arterial road is occupied with positive parked vehicles, i.e., $n_i^e(x) > 0$, $\forall x \in \left[0, \overline{x_i}\right)$ at spatial equilibrium, until it reaches the equilibrium parking span $x = \overline{x_i^e}$, $(i = a, c)$.*

**Proof.** This lemma can be proved by contradiction. Let $x^1 < x^2$, if there exists a parking gap $\left[x^1, x^2\right]$ in the linear corridor where no one in type $i(=a,c)$ chooses to park in between, it should satisfy $n_i(x^1) = 0 = n_i(x^2)$, $S_i(x^1) = S_{\min} = S_i(x^2)$.

From $P_i(x^1) = P_i(x^2)$, there is $x^1 = x^2$. In other words, there is no such parking gap.

□

**Lemma 3.2** *At unpriced equilibrium, if $n_i(x)$ is the only variable regarding $x$ in $S_i(x)$, the numbers of AV and HV travellers choosing to park at any location always decrease with the rising distance from CBD, i.e., $n_i^e{'}(x) < 0$ $(i = a, c)$.*

**Proof.** From $P_i{'}(x) = 0$, $(i = a, c)$,

$$S_a{'}(x) = \frac{-\lambda_a - \beta_a n_a^e(x)}{\gamma_a} < 0, \quad S_c{'}(x) = \frac{-\lambda_c + \beta_c n_c^e(x)}{\gamma_c}. \tag{7}$$

As $n_i(x)$ is the only variable in $S_i(x)$ changed with $x$, we have $S_i{'}(x) = \frac{\partial S_i(x)}{\partial n_i(x)} n_i^e{'}(x)$. From $\frac{\partial S_i(x)}{\partial n_i(x)} > 0$, there is $n_a^e{'}(x) < 0$. As for HVs, since we assume $\lambda_c > \beta_c k_{max} \geq \beta_c n_c^e(x)$, it always satisfies $n_c^e{'}(x) < 0$. □

From Lemma 3.1 and Lemma 3.2, one can verify that at the furthest parking location $x = \overline{x_i}$, there is

$$n_i(\overline{x_i}) = 0. \tag{8}$$

In addition, it is intuitive to find that, even though the directions of parking cruising for AVs and HVs are opposite, both of them prefer to park close to CBD if it satisfies that $n_i(x)$ is the only variable regarding $x$ in $S_i(x)$[3]. On one hand, if HVs park at locations with longer distance from CBD, they need to bear a much higher walking cost, despite lower cruising delay cost. HV travellers still prefer to park in city centre. On the other hand, for AVs, the further from CBD they park, the larger cruising cost they have and hence AVs will prefer to park close to the CBD in the case of unpriced spatial equilibrium.

---

[3] To note, this constraint of $S_i(x)$ guarantees that no other variables in $S_i(x)$ change with $x$. Nevertheless, if there is another variable regarding $x$ in $S_i(x)$, such as $m_i(x)$, Lemma 3.2 may not hold any more.

As it is the first work to incorporate parking cruising externalities in the environment of mixed AVs and HVs, we also compare the equilibrium with the benchmark case without considering parking cruising, i.e., $\beta_i = 0$ $(i = a, c)$ and have the following proposition.

**Proposition 3.1** *If parking cruising effects are ignored, when $n_i(x)$ is the only variable regarding $x$ in $S_i(x)$, the parking span is underestimated for HVs ( $\overline{x_c^e}\big|_{\beta_c>0} > \overline{x_c^e}\big|_{\beta_c=0}$ ) and is overestimated for AVs ( $\overline{x_a^e}\big|_{\beta_a>0} < \overline{x_a^e}\big|_{\beta_a=0}$ ) at unpriced equilibrium.*

**Proof.** Let $n_i^e(x)\big|_{\beta_i=0}$ and $n_i^e(x)\big|_{\beta_i>0}$ denote the parking location choices without and with cruising effects, respectively. Obviously, $\dfrac{\partial S_i(x)}{\partial n_i(x)\big|_{\beta_i=0}} = \dfrac{\partial S_i(x)}{\partial n_i(x)\big|_{\beta_i>0}} > 0$ holds. Since $n_i(x) \geq 0$, from Eq.(7), there is

$$S_a{}'(x)\big|_{\beta_a>0} = \frac{-\lambda_a - \beta_a n_a^e(x)}{\gamma_a} < \frac{-\lambda_a}{\gamma_a} = S_a{}'(x)\big|_{\beta_a=0} < 0 \quad \text{and}$$

$S_c{}'(x)\big|_{\beta_c>0} = \dfrac{-\lambda_c + \beta_c n_c^e(x)}{\gamma_c} > \dfrac{-\lambda_c}{\gamma_c} = S_c{}'(x)\big|_{\beta_c=0}$. For AVs, we can then derive that $n_a^e{}'(x)\big|_{\beta_i>0} < n_a^e{}'(x)\big|_{\beta_i=0} < 0$. That is to say, the number of parked AVs decreases at a faster rate with respect to $x$ when cruising effects are considered. With the conservation condition of travellers $N_a$ as in Eq.(2), we have $\overline{x_a^e}\big|_{\beta_a>0} < \overline{x_a^e}\big|_{\beta_a=0}$.

Similarly, for HVs, $0 > n_c^e{'}(x)\big|_{\beta_c>0} > n_c^e{'}(x)\big|_{\beta_c=0}$. While the number of parked HVs without considering cruising always decreases with $x$, the number of parked HVs decreases at an even slower rate when considering cruising. With the conservation of travellers $N_c$, we have $\overline{x_c^e}\big|_{\beta_c>0} > \overline{x_c^e}\big|_{\beta_c=0}$. □

In fact, for HVs, we have further generalized the result in (Anderson et al., 2004) where it was proved with specified $S_c(x)$ that larger $\beta_c$ leads to larger $\overline{x_c^e}$ at unpriced equilibrium. Intuitively speaking, without considering cruising effects, neither the trade-offs of HVs between walking and cruising, nor the inclinations of AVs to park closer to CBD to avoid additional self-cruising can be captured. If the spatial equilibrium solution of travellers' parking location choices is biased due to the ignorance of cruising effects, the resultant planning for parking space supply may deviate from the true optimal solution in practice, which further endorse the importance of incorporating cruising effects of HV and AV parking behaviours.

*3.1    Derivation of the spatial equilibrium*

In this subsection, we briefly present the derivation of the parking equilibrium with specified parking searching time functions.

Indeed, while the equilibrium is solved based on Eq.(2), (6) and (8), the general parking searching time function $S_i(x)$ impedes our determination of $n_i(x)$, $\overline{x_i}$ and $p_i$ as its interrelationship to $n_i(x)$ has not been revealed at length. Additionally, as the searching time is related to the parking supply, the specification of $S_i(x)$ also facilitates us in the exploration of optimal parking planning design in later discussion. For illustration purpose, the classic bi-nominal assumption-based parking searching time function (Anderson et al., 2004) satisfying Eq. (1) is next specified as $S_i(x) = \dfrac{m_i(x)}{m_i(x) - n_i(x)}$. Note that in our form of $S_i(x)$, not only $n_i(x)$ but also $m_i(x)$ change with $x$ and it can be verified that Lemma 3.2, Proposition 3.1 and later Proposition 4.1 hold when $m_i'(x) = 0$ (e.g., in the case with constant $\theta$ and $k$ along the city).

Thus, take AV as an example, after we substitute the specified $S_a(x)$ into $P_a'(x) = 0$, a partial differentiated equation with respect to $n_a(x)$ can be obtained for given $m_i(x)$,

$$n_a'(x) = \frac{-\left(m_a(x) - n_a(x)\right)^2 \left(\lambda_a + \beta_a n_a(x)\right)}{\gamma_a m_a(x)} + n_a(x) \frac{m_a'(x)}{m_a(x)}.$$

Together with $n_a(\overline{x_a}) = 0$ and $\int_0^{\overline{x_a}} n_a(x) dx = N_a$, the equilibrium $n_a^e(x)$ and $\overline{x_a^e}$ can be determined numerically (See Section 6 for numerical studies). Also, from Eq.(6), the equilibrium parking cost for AVs becomes,

$$p_a = P_a\left(\overline{x_a^e}\right) = \lambda_a \overline{x_a^e} + \beta_c N_c + \beta_a N_a + \gamma_a.$$

Likewise, for the derivation of optimal parking distribution discussed in Section 4, we just need to replace $P_a(x)$ with the marginal parking cost $MP_a(x)$ stated in (13) and the other steps remain the same as in the equilibrium derivation.

While the equilibrium and optimum can be barely determined analytically in this bi-nominal form of $S_i(x)$, we also make approximations to apply a piecewise linear form of $S_i(x)$ later in Section 5 to better excavate the properties of optimal parking planning design in mixed traffic.

4. **The optimum at lower level problem with given parking supply**

In the last section, we have scrutinized the travellers' equilibrium parking behaviours, wherein the individual parking cost is minimized. It is of our interest to turn to the question on the total parking cost minimization with managing measures, such as parking pricing and parking planning design. In light of the mixed AVs and HVs, both the travellers' parking behaviours and regulators' parking planning decisions are formulated and examined in a bi-level PPDP in the coming two sections.

With a predetermined $\{\theta(x), k(x)\}$, the LLP in the PPDP is first discussed in this section to encapsulate the optimal spatial parking location choices of travellers $n(x)$ with appropriate parking pricing. After that, in Section 5, from the perspective of regulators, the constraint on parking infrastructure expenditure is taken into account in the ULP to seek an optimal planning design minimizing the total parking cost.

To start with, we explore the LLP following Section 3 to determine the optimal spatial parking location choices for given $k(x)$ and $\theta(x)$. To solve the LLP optimum, we assume that the planners and regulators determine the optimal flows of travellers choosing to park at each location $n_i(x)$, such that the total parking cost of travellers is minimized. When solving the optimal solution, $n_a(x)$ and $n_c(x)$ are independent of each other with given $\{\theta(x), k(x)\}$. Therefore, the following minimization problem can be formulated as below (with the superscript $o$ indicating the optimum):

$$TP^{\min} = \min_{\{n_a(x)\}} \int_0^{\overline{x_a}^o} P_a(x) n_a(x) dx + \min_{\{n_c(x)\}} \int_0^{\overline{x_c}^o} P_c(x) n_c(x) dx, \qquad (9)$$
$$\text{s.t.} \int_0^{\overline{x_a}^o} n_a(x) dx = N_a, \int_0^{\overline{x_c}^o} n_c(x) dx = N_c.$$

Note that, in terms of total cruising cost, the following lemma is observed.

**Lemma 4.1** *The total cruising cost of both HVs and AVs $\int_0^{\overline{x_i}} n_i(x) c_i(x) dx$ are constant and independent of $n_i(x)$.*

**Proof.** From Eq.(3), the total cruising cost for HVs is $\beta_c \int_0^{\overline{x_c}} n_c(x) \int_x^{\overline{x_c}} n_c(u) du dx$. Let $F_c(x) = \int_x^{\overline{x_c}} n_c(u) du$ and hence $F_c'(x) = -n_c(x)$, $F_c(\overline{x_c}) = 0$ and $F_c(0) = N_c$. Obviously, the cost becomes $-\beta_c \int_0^{\overline{x_c}} F_c'(x) F_c(x) dx = -\frac{\beta_c}{2}\left(F_c^2(\overline{x_c}) - F_c^2(0)\right)$ $= \frac{1}{2}\beta_c N_c^2 = \frac{1}{2}\beta_c (1-\varepsilon)^2 N^2$, which is independent of the parking allocation $n_c(x)$. This constant total HV cruising cost has also been proved in (Anderson et al., 2004). Likewise, from Eq. (4), for the total AV cruising cost $\int_0^{\overline{x_a}} n_a(x) \left(\beta_c N_c + \beta_a \int_0^x n_a(u) du\right) dx$, we let $F_a(x) = \int_0^x n_a(u) du$ and there are $F_a'(x) = n_a(x)$, $F_a(\overline{x_a}) = N_a$ and $F_a(0) = 0$. The cost becomes

$$\beta_c N_c \int_0^{\overline{x_a}} n_a(x) dx + \beta_a \int_0^{\overline{x_a}} F_a'(x) F_a(x) dx = \beta_c N_c N_a + \frac{1}{2}\beta_a N_a^2$$

$= \beta_c (1-\varepsilon)\varepsilon N^2 + \frac{1}{2}\beta_a \varepsilon^2 N^2$, which does not change with $n_a(x)$ neither. □

From Lemma 4.1, the total cruising cost of all vehicles is also constant. In addition, Lemma 4.1 suggests that the solutions to the minimization problem (9) remain unchanged even if we subtract all the costs related to parking cruising. Therefore, let $TP_i = \int_0^{\overline{x_i^o}} \left(P_i(x) - c_i(x)\right) n_i(x) dx$, $i(=a,c)$, and the minimization problem (9) can be reduced to

$$TP^{\min} = \min_{\{n_a(x)\}} TP_a + \min_{\{n_c(x)\}} TP_c + TCr, \tag{10}$$

where $TCr = \frac{1}{2}\left(\beta_c(1-\varepsilon^2) + \beta_a \varepsilon^2\right) N^2$ to denote the total cruising cost and the other constraints are the same as in (9).

Further, it can be verified that for any $u \in \left[0, \overline{x_i^o}\right]$ $(i = a, c)$,

$$\frac{\partial^2 TP_i}{\partial n_i^2(u)} = \gamma_i \left( 2\frac{\partial S_i(u)}{\partial n_i(u)} + \frac{\partial^2 S_i(u)}{\partial n_i^2(u)} n_i(u) \right) > 0, \quad \frac{\partial^2 TP_i}{\partial n_i(u) \partial n_i(x \neq u)} = 0. \quad (11)$$

That is to say, $TP_a$ and $TP_c$ have unique minimums in terms of parking distribution $n_i(x)$. Consequently, similar to the system optimum in transportation networks (Sheffi, 1985), in order to obtain the solution $\{n_i^o(x), \overline{x_i^o}\}$ to minimization problem (10), it suffices to equate the marginal parking cost (with respect to $n_i(x)$) for all locations with positive parking (Anderson et al., 2004), which satisfies

$$MP_i'(x) = 0 \quad \text{for} \quad x \in \left[0, \overline{x_i^o}\right], \quad (i = a, c), \quad (12)$$

where $MP_i(x) = \dfrac{\partial((P_i(x) - c_i(x))n_i(x))}{\partial n_i(x)} = \lambda_i x + \gamma_i \left( S_i(x) + n_i(x)\dfrac{\partial S_i(x)}{\partial n_i(x)} \right). \quad (13)$

Indeed, $MP_i(x)$ can be interpreted as the marginal contribution of an additional traveller parking at $x$ to the total parking cost of vehicles in the same type at this location, excluding the cruising cost $c_i(x)$.

Regarding the LLP optimum, the trend of optimal parking distribution is observed.

**Lemma 4.2** *At the LLP optimum with given* $\{\theta(x), k(x)\}$, *if* $n_i(x)$ *is the only variable regarding* $x$ *in* $S_i(x)$, *the number of both AV and HV parkers decreases with* $x$, *i.e.,* $n_i^o{}'(x) < 0, \ (i = a, c).$

**Proof.** From Eq.(12), there is

$$n_i^o{}'(x) = -\frac{\lambda_i}{\gamma_i} \left( 2\frac{\partial S_i(x)}{\partial n_i(x)} + n_i^o(x)\frac{\partial^2 S_i(x)}{\partial n_i^2(x)} \right)^{-1} < 0. \quad \square \quad (14)$$

Comparing Lemma 3.2 and Lemma 4.2, we further identify the parking span difference of AVs between unpriced equilibrium and LLP optimum, as summarized in the following proposition.

**Proposition 4.1** *With given* $\{\theta(x), k(x)\}$, *if* $n_i(x)$ *is the only variable regarding* $x$ *in* $S_i(x)$, *the parking span of AVs at LLP optimum is always no less than that at unpriced equilibrium, i.e.,* $\overline{x_a^e} \leq \overline{x_a^o}$.

**Proof.** Let $f(n_a) = \dfrac{\partial S_a(x)}{\partial n_a(x)} > 0$. With the chain rules and Eq.(7), there is

$$f(n_a^e) = \frac{-\lambda_a - \beta_a n_a^e(x)}{\gamma_a n_a^{e'}(x)}, \text{ and thus } n_a^{e'}(x) = -\frac{\lambda_a + \beta_a n_a^e(x)}{\gamma_a} \frac{1}{f(n_a^e)}.$$ 

From Eq.(14), $n_a^{o'}(x) = -\dfrac{\lambda_a}{\gamma_a}\left(\dfrac{1}{2f(n_a^o) + n_a^o(x) f'(n_a^o)}\right)$. Additionally, since $f'(n_a) = \dfrac{\partial^2 S_a(x)}{\partial n_a^2(x)} \geq 0$, larger $n_a$ leads to larger $f(n_a)$. It can then be verified readily that $0 > n_a^{o'}(x) \geq n_a^{e'}(x)$ for all $x$ with $n_a^o(x) \geq n_a^e(x)$. As $n_a^e(x)$ and $n_a^o(x)$ are decreasing with $x$ and their positive integrals over $x$ remain a constant $N_a$, they must intersect once at least, with an interval where $n_a^o(x) \geq n_a^e(x)$. In this interval, the number of parked vehicles will decrease at a slower or equal rate in LLP optimum than in equilibrium and it is impossible for $n_a^e(x)$ to exceed $n_a^o(x)$ any more. Therefore, it can be concluded that $\overline{x_a^e} \leq \overline{x_a^o}$.
□

The intuition underlying Proposition 4.1 is the uninternalized externalities at unpriced equilibrium. At equilibrium, AV travellers prefer to park close to CBD to lower his/her cost of cruising and self-driving without caring about the extra searching costs imposed upon other vehicles, which inevitably leads to higher parking searching cost at the locations near CBD. However, at optimum, as AVs' total cruising cost on the arterial road is a constant, a dispersed parking distribution of AV reduces the total searching cost and hence the AV parking span at optimum is wider. Indeed, without considering parking cruising, the similar property of AV parking span has also been discovered in (Liu, 2018), where the time-dependent congestion is taken into account.

Nevertheless, for HVs, the change in parking span from equilibrium to optimum is ambiguous. It is worth noting that in (Anderson et al., 2004), it was concluded that the optimum involves less tight parking for HVs without considering cruising effects. However, this seems to be no longer the case, when parking cruising is considered. In fact, such a comparison of HV has never been made in (Anderson et al., 2004)'s later discussions. With parking cruising, there is already less crowding with a wider parking span at equilibrium, as shown in Proposition 3.1. The willingness to park more inward has been counteracted by cruising. As the total cruising cost of HVs stays the same and does not affect the LLP optimum, it remains unclear whether the parking distribution will be further dispersed at optimum.

*4.1     Optimal differentiated parking pricing*

To internalize the parking externalities of both types of vehicles, we also determine the optimal differentiated parking price $\tau_i(x)$ $(i=a,c)$ as the difference between marginal parking cost and the generalized parking cost as follows,

$$\tau_a(x) = \beta_a \int_x^{\overline{x_a^o}} n_a^o(u) du + \gamma_a n_a^o(x) \frac{\partial S_a(x)}{\partial n_a^o(x)}, \tag{15}$$

$$\tau_c(x) = -\beta_c \int_x^{\overline{x_c^o}} n_c^o(u) du + \gamma_c n_c^o(x) \frac{\partial S_c(x)}{\partial n_c^o(x)}. \tag{16}$$

One may notice that the optimal parking price is independent of HV walking cost and AV self-driving cost. On the one hand, both prices in Eq.(15),(16) internalize the searching externalities imposed on all vehicles at that location. On the other hand, they also internalize the cruising externalities imposed on all AVs parked at more outward locations and on all HVs parked at more inward locations from CBD. Moreover, for AVs, it can easily be verified that $\frac{\partial \tau_a(x)}{\partial x} < 0$ and hence the optimal parking price of AVs is the highest at CBD and decreases with $x$. However, for HVs, it cannot be determined whether the optimal parking price decreases with $x$ without specifying $S_i(x)$, when cruising is considered.

In addition, let $TC$ denote travellers' total parking cost, including optimal parking pricing at LLP optimum. It can be represented by the sum of total marginal parking cost and total cruising cost,

$$TC = MP_a N_a + MP_c N_c + TCr, \qquad (17)$$

which is different from $TP^{\min}$ in (10). In fact, one may verify that $TC - TP^{\min}$ represents the total net expense of parking pricing for travellers.

Besides, even though the optimal parking distribution can be achieved by differentiated pricing, it should be noted that $\{\theta(x), k(x)\}$ may hinder the system from reaching the minimum total parking cost. For instance, an excess supply of parking spaces to AVs would squeeze the space for HVs such that they need to park much further and walk more with higher parking cost, even optimal pricing is implemented. Thus, it is of immense significance to unearth the optimal parking planning design for AVs and HVs, which will be discussed in the next section.

So far, we have explored the properties at the LLP optimum in (9) with given $\{\theta(x), k(x)\}$ analytically, without specifying the parking searching time function.

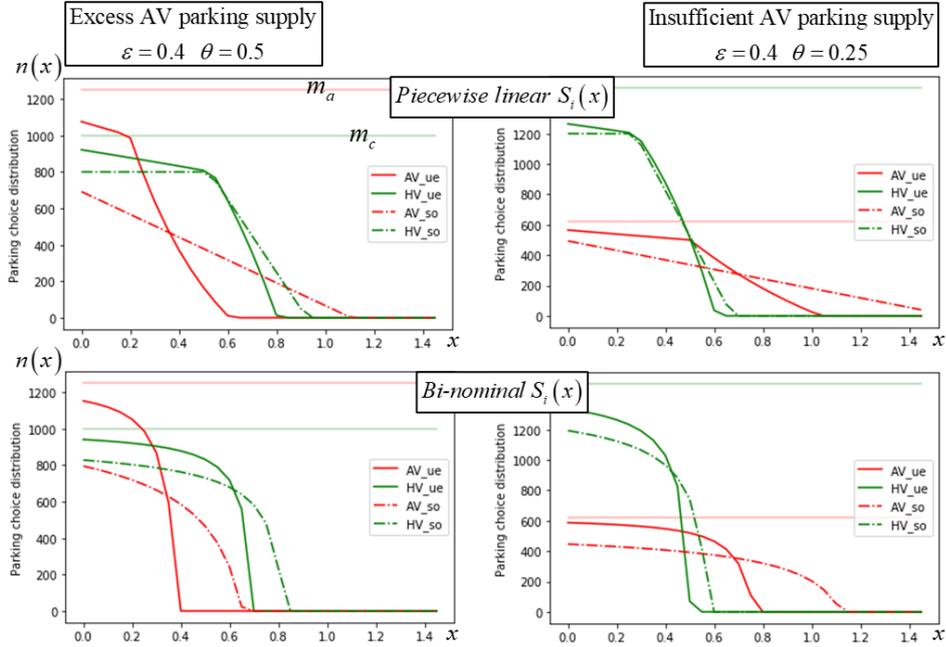

Figure 2 The parking distribution along the city at unpriced equilibrium and at optimum with given parking supply. (Numerical examples)

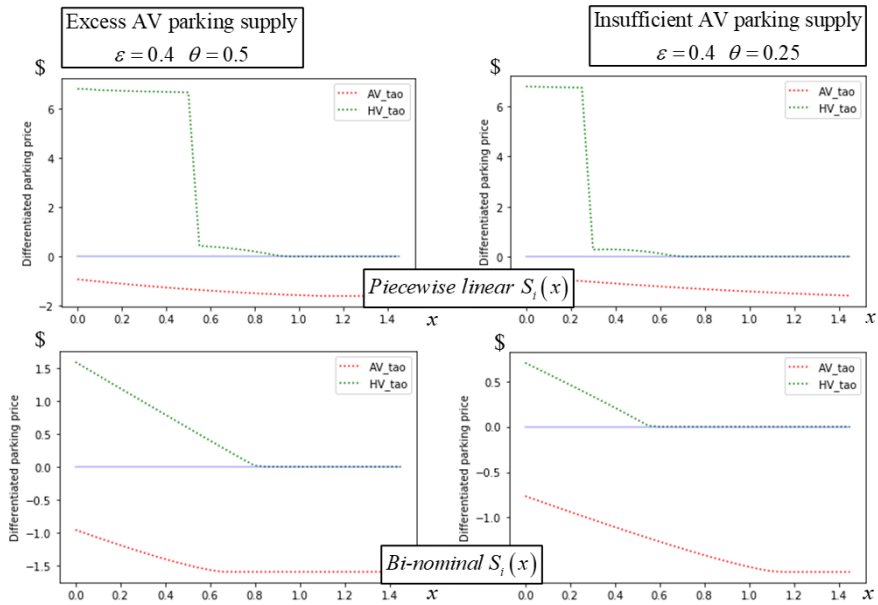

Figure 3 The optimal differentiated parking pricing along the city with given parking supply. (Numerical examples)

## 5. Upper level optimum

We next investigate ULP in the PPDP and aim to minimize the total parking cost of travellers. It is postulated that the total parking cost including parking pricing has reached the corresponding minimum $TC$ [4] for the given parking supply as in Section 4. Therefore, the ULP is then formulated as,

$$\min_{\{\theta(x), k(x)\}} TC,$$
$$\text{s.t. } \theta(x) \in [0,1], k(x) > 0, \forall x \in [0, x] \quad (18)$$

$$NP \leq \overline{NP}, \quad (19)$$

$$\text{and } \int_0^x m_a(u)du > N_a, \int_0^x m_c(u)du > N_c, \quad (20)$$

where $x$ denotes supply span of parking spaces assuming $\overline{x_a}, \overline{x_c} < x$ and $NP$ (and $\overline{NP}$) is the actual (and budget) aggregate infrastructure cost of parking spaces along the city. Here, Eq.(19) and (20) represent the constraints on infrastructure cost and parking supply respectively. Recall that, the user reaction in this ULP for each parking planning design is the parking location choice distribution $n(x)$ along the city.

In fact, in view of the parking demand of mixed AVs and HVs, regulators need to invest in adjusting and upgrading the parking spaces for dedicated AV usage. At the same time, it is also anticipated that the total parking land rent can remain unchanged or even be reduced with optimized land supply in the presence of AVs, as AV travellers can park further away at the parking spaces with a cheaper land rent. On this account, $NP$ is defined as the sum of total upgrade cost for AV parking and the total parking land rent, i.e.,

$$NP = v_a \int_0^x \theta(u)k(u)du + \int_0^x L(u)k(u)du, \quad (21)$$

where $v_a$ denotes the upgrade cost per unit area for AV parking and $L(x)$ represents the parking land rent at $x$, which is assumed to remain unchanged over time in our investigation period.

---

[4] Though the expense of parking pricing can be collected by regulators as social revenue, it is still part of the parking cost perceived by travellers, which we aim to minimize. Thus, we adopt $TC$ in the ULP formulation instead of $TP^{\min}$.

Up till now, we have defined all the terms of ULP minimization in PPDP. Admittedly, it is of considerable difficulty to obtain its solution $\{\theta^o(x), k^o(x)\}$ with the current formulation. As mentioned in Section 3.1, without specifying the searching time function, parking location choices as well as the LLP minimum can hardly be determined. In fact, just like most other NDP, the proposed PPDP is a NP-hard problem and the global optimality cannot be guaranteed due to the non-convexity in the bi-level programs (Farahani et al., 2013; Levin et al., 2020).

Nevertheless, some indicative properties of the ULP optimum can still be captured analytically, with the following specifications on the searching time function and parking planning.

First, we apply a piecewise linear parking searching time function as follows.

$$S_i(x) = \begin{cases} \delta_i \dfrac{n_i(x)}{m_i(x)}, & \text{if } \dfrac{n_i(x)}{m_i(x)} \leq 1-\omega_i; \\ \delta_i(1-\omega_i) + \Delta_i\left(\dfrac{n_i(x)}{m_i(x)} - 1 + \omega_i\right), & \text{if } 1-\omega_i < \dfrac{n_i(x)}{m_i(x)} \leq 1; \end{cases} \quad (22)$$

where $\delta_i$ is a constant positive coefficient which links the effective occupancy of parking space to parking searching time, $\omega_i$ is a sufficiently small positive number close to 0 to indicate the proportion of vehicles with a dramatic searching time increase, and $\Delta_i$ is a large coefficient approaching infinity (Qian et al., 2014; Su & Wang, 2019). We further let $\delta_a = \delta_c = \delta$, $\omega_a = \omega_c = \omega$ and $\Delta_a = \Delta_c = \Delta$ for simplicity, and hence the difference of parking searching time of AVs and HVs lies in the allocated number of parking spaces $m_i(x)$ and the actual parking distribution $n_i(x)$ only.

As for the parking planning, we hereinafter assume the parking areas per unit distance (or total parking width) is constant along the city, and so is the proportion of AV parking areas for analytical tractability, by letting $k(x) = k$ and $\theta(x) = \theta$. While the constant parking areas per unit distance $k$ assumption can often be found in literature (Arnott et al., 1991; Anderson et al., 2004; Liu, 2018), we further assume consistent $\theta$ along the city to reduce the complexity of the algebra. One may also refer $\theta$ as the expected parking area allocation for AVs along the city. Thus, $m_a = \dfrac{\theta k}{\phi}$ and $m_c = (1-\theta)k$ are constant along the city.

Indeed, with the specifications mentioned above, following Section 3.1, the spatial equilibrium and optimum in the LLP at given parking supply can now be approximately derived with the parking spans and parking cost summarized in Table 1 and numerical results presented in the upper halves of Figure 2 and Figure 3. With constant $m_a, m_c$, $n(x)$ is the only variable regarding $x$ in $S(x)$ and one can verify the properties presented in previous sections.

Table 1  Summary of LLP equilibrium and optimum with piecewise linear searching time function.

| | Unpriced Spatial Equilibrium | Optimum |
|---|---|---|
| AV | $\overline{x_a^e} = \dfrac{N_a}{m_a(1-\omega)} - \dfrac{\gamma_a \delta}{m_a \beta_a} - \dfrac{\gamma_a \delta(\lambda_a + m_a \beta_a(1-\omega))}{m_a^2 \beta_a^2 (1-\omega)} \ln\left(\dfrac{\lambda_a}{\lambda_a + m_a \beta_a (1-\omega)}\right)$ | $\overline{x_a^o} = \dfrac{N_a}{m_a(1-\omega)} + \dfrac{\gamma_a \delta(1-\omega)}{\lambda_a}$ |

| | | | | |
|---|---|---|---|---|
| | | $p_a = \beta_c N_c + \beta_a N_a + \lambda_a \overline{x_a^e}$ | | $MP_a = \lambda_a \overline{x_a^o}$ |
| HV | $\overline{x_c^e} = \frac{N_c}{m_c(1-\omega)} + \frac{\gamma_c \delta}{m_c \beta_c} - \frac{\gamma_c \delta(\lambda_c - m_c \beta_c (1-\omega))}{m_c^2 \beta_c^2 (1-\omega)} \ln \frac{\lambda_c}{\lambda_c - m_c \beta_c (1-\omega)}$ | | $\overline{x_c^o} = \frac{N_c}{m_c(1-\omega)} + \frac{\gamma_c \delta(1-\omega)}{\lambda_c}$ | |
| | $p_c = \lambda_c \overline{x_c^e}$ | | $MP_c = \lambda_c \overline{x_c^o}$ | |

Moreover, the parking infrastructure cost in (21) can be reduced to $NP = k(v_a \theta x + L)$ with $L = \int_0^x L(u) du$. Thus, from Eq.(17) and Table 1, the minimization problem in (18) becomes

$$\min_{\{\theta,k\}} TC = C(\varepsilon) + \frac{N^2}{k(1-\omega)} \left( \frac{\lambda_c(1-\varepsilon)^2}{(1-\theta)} + \phi \frac{\lambda_a \varepsilon^2}{\theta} \right),$$
$$\text{s.t. } \theta \in [0,1], k > 0,$$
$$k(v_a \theta x + L) \leq \overline{NP},$$
$$\text{and } (20), \quad (23)$$

where $C(\varepsilon) = (\gamma_c(1-\varepsilon) + \gamma_a \varepsilon) \delta(1-\omega) N + \frac{1}{2}(\beta_c(1-\varepsilon^2) + \beta_a \varepsilon^2) N^2 > 0.$

Note that in (23) we distinguish the terms related to parking planning design $\{\theta, k\}$ from the others. These other terms independent of $\{\theta, k\}$ are related to parking searching and cruising, and are further summarized as a function of $\varepsilon$, $C(\varepsilon)$. In fact, due to the internalized externalities at optimum, the term related to parking searching $\{\gamma_a, \gamma_c\}$ in total parking cost $TC$ no longer depends on the parking planning design $\{\theta, k\}$, but it is a function of the critical parking occupancy $(1-\omega)$ instead, which is a constant in our model. In addition, the total cruising cost does not vary with $\{\theta, k\}$ neither.

For later contrast and comparison, we further define the benchmark case (denoted by $b$) as the situation with a given initial parking area (width) $k^b > 0$ and the parking areas are allocated proportional to the required areas at different AV penetration $\varepsilon$, i.e.,

$$\theta^b = \frac{\varepsilon\phi}{\varepsilon\phi + (1-\varepsilon)} \leq \varepsilon. \tag{24}$$

Further, the parking budget $\overline{NP}$ in (23) is calculated as the parking infrastructure cost in this benchmark case, i.e.,

$$\overline{NP} = k^b \left( v_a \theta^b x + L \right), \tag{25}$$

which is the budget amount to guarantee sufficient proportion of parking supply for both AV and HV parking.

Consider $\{\theta, k\}$ as the decision variables of $TC$ in (23), we now revisit the ULP first at the first-best optimum in Section 5.1 and then considering the additional upper bound of $k$ at second-best optimum in Section 5.2.

*5.1 First-best optimum*

First and foremost, the first-best optimum (denoted by $o1$) for minimization problem in (23) is explored and the following proposition and corollaries are observed.

**Proposition 5.1** *For every given AV penetration $\varepsilon$, the first-best optimal parking planning design to minimize travellers' total parking cost at optimal pricing is,*

$$\theta^{o1} = \frac{\sqrt{\mu\phi}\varepsilon}{\sqrt{\mu\phi}\varepsilon + (1-\varepsilon)\sqrt{\frac{L+v_a x}{L}}} \leq \varepsilon, \quad k^{o1} = \frac{\overline{NP}}{v_a \theta^{o1} x + L}. \tag{26}$$

*where $\mu = \lambda_a / \lambda_c < 1$ and the equality for $\theta^{o1}$ holds if and only if $\varepsilon = 0, 1$.*

**Corollary 5.1** When $\lambda_a / \lambda_c < \frac{L + v_a x}{L}\phi$, there is $\theta^{o1} \leq \theta^b$ and hence $k^{o1} \geq k^b$, vice versa.

**Corollary 5.2** $\theta^{o1}$ does not change with $k$, decreases with $v_a$, but increases with $\varepsilon$, $\phi$, $L$ and also increases with the unit distance cost ratio between self-driving and walking.

**Proof.** From Eq.(23), there is $k \leq \frac{\overline{NP}}{v_a \theta x + L}$ and hence

$$TC \geq C(\varepsilon) + \frac{N^2}{(1-\omega)\overline{NP}}\left(\frac{\lambda_c(1-\varepsilon)^2}{(1-\theta)} + \phi\frac{\lambda_a \varepsilon^2}{\theta}\right)(v_a \theta x + L)$$ and the equality holds if

and only if $k$ reaches maximum at $\frac{\overline{NP}}{v_a \theta x + L}$. Therefore, the minimization problem in (23) can be reduced to find $\min_\theta \left(\frac{\lambda_c(1-\varepsilon)^2}{(1-\theta)} + \phi\frac{\lambda_a \varepsilon^2}{\theta}\right)(v_a \theta x + L)$. It can be verified that when $\theta < \theta^{o1}$, the total parking cost at optimal pricing decreases with $\theta$ and vice versa. Total parking cost reaches its minimum if and only if $\theta = \theta^{o1}$ and

$$\min\left(\frac{\lambda_c(1-\varepsilon)^2}{(1-\theta)} + \phi\frac{\lambda_a \varepsilon^2}{\theta}\right)(v_a \theta x + L) = \lambda_c\left(\sqrt{L}(1-\varepsilon) - \sqrt{v_a x + L}\sqrt{\mu\phi}\varepsilon\right)^2.$$ Once $\theta^{o1}$

is obtained, there is $k^{o1} = \frac{\overline{NP}}{v_a \theta^{o1} x + L}$ and Proposition 5.1 is proved. From Eq.(25),

(26) and the upper bounded infrastructure cost, Corollary 5.1 and 5.2 can also be derived. □

As shown in Proposition 5.1, it can be verified that $\theta^{o1} \in [0,1]$ and $\theta^{o1}$=0,1 if and only if $\varepsilon$=0,1. Indeed, for the extreme cases with all AVs or all HVs, there is no need to allocate parking spaces for other types of vehicles. However, for the mixed case with both AVs and HVs, though the more AVs leads to the larger $\theta^{o1}$, our result suggests that this optimal area proportion should be smaller than the given AVs' penetration, i.e., $\theta^{o1} \leq \varepsilon$. Meanwhile, the optimal parking land area $k^{o1}$ reaches the maximum for the corresponding $\theta^{o1}$ at the given infrastructure budget constraint.

In fact, with the state-of-the-art technology of self-driving, the AV self-driving cost at unit distance can be much smaller than the walking cost for HV drivers. Namely, the inequality $\lambda_a / \lambda_c < \frac{L+v_a x}{L}\phi$ generally holds. Hence, Corollary 5.1**Error! Reference source not found.** implies that at optimum, the parking area allocated to AVs can be even smaller than average areas required to provide enough parking spaces for AV parking at penetration $\varepsilon$. In addition, with the budget constraint, the smaller than required $\theta$ also implies the larger $k$ at the first-best optimum. Admittedly, expanding the total parking land area effectively releases the tension in parking supply and therefore lowers the total parking cost enormously. From the perspective of regulators, Corollary 5.1 suggests that for any given AV penetration with limited budget on parking infrastructure, rather than using the budget up to upgrade the parking area required for AV with $\theta = \theta^b$, it is more cost beneficial to spend some on the enlargement of parking land area, if possible.

Moreover, Corollary 5.2**Error! Reference source not found.** finds that the $\theta^{o1}$ is independent of $k$. That is to say, for arbitrary constant parking land area supply, the optimal parking area allocation to minimize total parking cost remains unchanged. On one hand, in terms of individual parking configuration, it is further suggested in Corollary 5.2**Error! Reference source not found.** that when the size of an AV parking space is reduced, or when the self-driving technology is getting more developed with even lower unit distance self-driving cost, it is beneficial to encourage AVs to cruise more distantly by further reducing their allocated proportion of parking areas. It is also noteworthy that due to the concise form of $\theta^{o_1}$, there is no need to identify the exact cost of walking or self-driving, but only their ratio is enough in the calculation of $\theta^{o_1}$. On the other hand, the parking infrastructure cost also affects the value of $\theta^{o1}$. Indeed, the high expense in AV parking upgrade and the low land rent will erode the advantages of AV in total parking cost minimization. Therefore, with higher AV upgrade cost (larger $v_a$), or with reduced aggregate land rent (smaller $L$), the optimal proportion $\theta^{o1}$ becomes smaller.

*5.2  Second-best optimum*

We have scrutinized the first-best optimum in ULP in last subsection, where the first-best optimal parking land area is usually larger than the original benchmark case. Nevertheless, in most metropolitans there is usually scarce land supply and it is extremely difficult to expand the parking land area along the city. If that is the case (denoted by $o2$), let say the parking land area is upper bounded by the initial area $k \leq k^b$, we have the following proposition on the second-best optimum.

**Proposition 5.2** *When the parking land area per unit distance is further upper bounded with $k \leq k^b$, the second-best optimum in ULP satisfies*

$$\theta^{o2} = \frac{\sqrt{\mu\phi\varepsilon}}{\sqrt{\mu\phi\varepsilon}+1-\varepsilon} \in \left[\theta^{o1}, \theta^b\right], \quad k^{o2} = k^b \qquad (27)$$

**Proof.** The proof of this proposition is similar to that of Proposition 5.1. From Corollary 5.1, we notice that at first-best optimum, $k^{o1} \geq k^b$ holds if $\lambda_a/\lambda_c < \frac{L+v_a x}{L}\phi$. Therefore, now after adding the upper bound of $k$, such constraint is usually binding and $k$ reaches its maximum to minimize $TTP_{\min}$ with $k^{o2} = k^b$. In other words, the optimal $k$ hardly change with $\theta$ and the minimization problem in (23) is reduced to find $\min_{\theta} \left( \frac{\lambda_c (1-\varepsilon)^2}{(1-\theta)} + \phi \frac{\lambda_a \varepsilon^2}{\theta} \right)$. It can be verified that the minimum is obtained at $\theta = \theta^{o2}$, and

$$\min \left( \frac{\lambda_c (1-\varepsilon)^2}{(1-\theta)} + \phi \frac{\lambda_a \varepsilon^2}{\theta} \right) = \lambda_c \left( (1-\varepsilon) - \sqrt{\mu \phi} \varepsilon \right)^2. \square$$

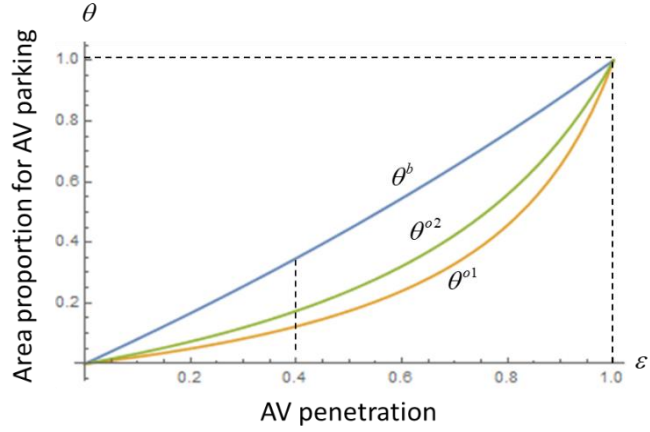

Figure 4 The first- and second-best optimal area allocation for AV parking to minimize travellers' total parking cost at different penetration of AVs.

From Proposition 5.1 and 5.2, one can notice that the second-best $\theta^{o2}$ is indeed the upper bound of the first-best $\theta^{o1}$ and it can be easily derived as well that $\theta^{o2} \leq \theta^b$ if $\lambda_a / \lambda_c < \phi$. Without the expansion of parking area at the second-best optimum, the area proportion for AV parking cannot be too small, otherwise the increase in the parking cost of AV will dominate in total parking cost of travellers, compared to the decrease in the cost of HV. In Figure 4, the AV parking area proportions at benchmark case, first- and second-best optimum are delineated at different AV penetration.

In fact, combining the results in Section 5.1 and 5.2, it can be found that some properties at first-best optimum can also apply to the second-best optimum. In line with Corollary 5.2, the second-best $\theta^{o2}$ neither changes with $k$ but increases with $\varepsilon$, $\phi$ and $\mu$. Furthermore, in both first- and second-best optimum, the optimal proportion of parking land area for AV is always no larger than that proportional to AVs' required parking area. Resulted from such smaller proportion, the insufficient parking supply for AVs in the land area or in the number of spots seems to cause inconvenience to AV travellers. Nonetheless, it should be pointed out that such insufficiency forces AVs to self-drive further outwards to park and meanwhile allows HVs to park closer to the city centre with less walking, which in turn lowers total parking cost in the system. This finding is indicative for regulators to design the parking supply of AVs and HVs.

To note, for the scenarios with $v_a > 0$, the equality in $\theta^{o2} \geq \theta^{o1}$ holds if and only if $\varepsilon = 0,1$, wherein $\theta^{o2} = \theta^{o1} = \theta^b = \varepsilon$. Interestingly, when we consider the case where there is negligible upgrade cost for AV parking compared to land rent, i.e., $v_a \to 0$, the minimization problem in (23) for first-best optimum can be reduced to find $\min_{\theta} \left( \frac{\lambda_c (1-\varepsilon)^2}{(1-\theta)} + \phi \frac{\lambda_a \varepsilon^2}{\theta} \right)$. This is the same as in the proof in Proposition 5.2, and hence $\theta^{o1} \big|_{v_a \to 0} = \theta^{o2}$. Moreover, as the infrastructure budget now depends on $k$ only and the larger $k$ leads to lower total parking cost, there is no incentive to reduce the parking land area and thus $k^{o1} \big|_{v_a \to 0} = k^b$ due to the budget constraint. The following lemma is concluded.

**Lemma 5.1** *When the upgrade cost for AV parking is negligible $(v_a \to 0)$, the first-best optimum in ULP is the same as the second-best optimum as shown in (27), i.e.,*

$$\theta^{o1} \big|_{v_a \to 0} = \theta^{o2}, \quad k^{o1} \big|_{v_a \to 0} = k^b. \tag{28}$$

It should be noted that from Proposition 5.2, there is no need to obtain the information on parking infrastructure cost in the determination of parking planning design at second-best optimum. Namely, in the situation with limited information on the parking land rent and upgrade cost for AV parking, regulators can turn to the second-best optimum design in total parking cost minimization. As shown in Lemma 5.1, such second-best result can reach the first-best optimum for negligible upgrade cost, regardless of the land rent.

*5.3    System performance*

So far, we have explored the properties at first- and second-best optimum in ULP. Naturally, it is of our interest to investigate the system performance of the proposed optimal parking planning design. We next compare and contrast the aforementioned three cases: benchmark case, the first-best optimum and the second-best optimum. The following lemmas hold.

**Lemma 5.2** *The reduction percentage of total parking cost from benchmark to first-best optimum satisfies* $\frac{TC^b - TC^{o1}}{TC^b} < \left( \frac{\sqrt{v_a x + L}\sqrt{\phi} - \sqrt{L}\sqrt{\mu}}{\sqrt{v_a x + L}\sqrt{\phi} + \sqrt{L}\sqrt{\mu}} \right)^2$.

**Lemma 5.3** *The reduction percentage of total parking cost from benchmark to second-best optimum satisfies* $\frac{TC^b - TC^{o2}}{TC^b} < \left( \frac{\sqrt{\phi} - \sqrt{\mu}}{\sqrt{\phi} + \sqrt{\mu}} \right)^2$.

**Lemma 5.4** *The reduction percentage of total parking cost from second-best optimum to first-best optimum satisfies* $\frac{TC^{o2} - TC^{o1}}{TC^{o2}} \leq \frac{v_a x \theta^b}{L + v_a x \theta^b}$, *and the equality holds if and only if* $v_a \to 0$.

**Proof of Lemma 5.2, Lemma 5.3 and Lemma 5.4.** See Appendix B. □

In Lemma 5.2 and Lemma 5.3, comparisons are made on the performance improvement from the benchmark case to the system optimums. One can readily verify that the total parking cost reduction in first-best optimum is larger than that in second-best optimum. Though Lemma 5.2 and Lemma 5.3 imply that the cost reduction is modest, it should be noted that such comparisons are conducted between the optimized cases, where the optimal differentiated pricing has been implemented. While the optimal pricing has already lowered the total parking cost, the proposed optimal parking planning design can further reduce the cost. Besides, we consider the benchmark case with matching area proportion of AV parking in our study, which has already lowered the total parking cost of commuters. If the system performances are to compare with the initial case without AV parking spaces, the cost reduction at system optimums will become much more remarkable.

In addition, Lemma 5.4 compares system performance between the first- and second-best optimum. From the proof of Lemma 5.4, We notice that the more advances in self-driving technology (smaller $\mu$), the larger performance improvement there will be from second-best to first-best optimum. This is because compared to second-best optimum, the first-best optimum encourages saving more upgrade cost for parking area expansion, resulting in smaller area proportions for AV parking at optimum. With improved self-driving technology, the AV parking cost further decreases and hence the effects of smaller $\theta$ (i.e., larger area proportions for HV parking) become more prominent in total parking cost minimization, which enlarges the performance improvement. Nonetheless, from Lemma 5.4, it can be observed that such performance improvement is upper bounded. If the proportion of upgrade cost in aggregate infrastructure cost drops, the performance improvement becomes smaller. In the extreme case when upgrade cost is negligible, the performance improvement becomes zero, which is in line with our result in Lemma 5.1.

In fact, for the second-best optimum, though the total parking cost reduction is smaller compared to that in first-best optimum, it also saves the infrastructure cost with lower parking budget spending, which is summarized below in Corollary 5.3.

**Corollary 5.3** *Unlike the first-best optimum fully utilize the budget on parking infrastructure, the second-best optimum saves the budget by the amount of* $k^b v_a \left( \theta^b - \theta^{o2} \right) x \in \left[ 0, k^b v_a \theta^b x \right)$.

From Corollary 5.3, the smaller $\theta^{o2}$ leads to larger cost saving on parking infrastructure. While the parking budget is not included in our objective function in the ULP, it is noteworthy that when regulators also aim to lower the parking budget, they should revisit their objective function. If that is the case, the second-best optimum in our study should outperform the first-best optimum.

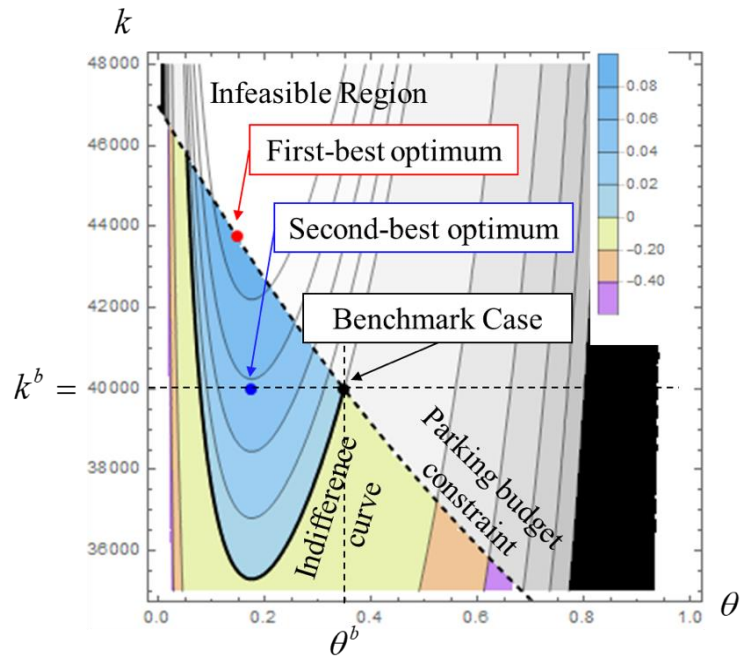

Figure 5 The contour plot of reduction percentage of total parking cost, with respect to different parking planning design $\{\theta, k\}$ (Numerical example).

## 6. Numerical example

Numerical examples are presented in this section. First, the case with constant $\{\theta, k\}$ is exemplified to illustrate and contrast the spatial parking equilibrium and LLP optimum in both bi-nominal and piecewise-linear searching time functions. After that, we exemplify the ULP optimum in the bi-level parking planning design problem to demonstrate the findings in Section 5. Lastly, we further relax the constant $\{\theta, k\}$ to reveal the parking equilibrium with location-dependent parking supply $\{\theta(x), k(x)\}$.

The related parameters are assigned as follows. Let $N = 20000$ (drivers), $\varepsilon = 0.4$, $\phi = 0.8$ (area/parking space). Apparently, it can be derived that $\theta^b = 0.35$ at the benchmark case. Also, the same parameters in (Anderson et al., 2004) are applied for HVs, i.e., $\lambda_c = 4$ (SGD/km), $\beta_c = 1 \times 10^{-4}$ (SGD), $\gamma_c = 0.1$ (SGD). As for AVs, we let $\lambda_a = 0.5$ (SGD/km), $\beta_a = 0.5 \times 10^{-4}$ (SGD), $\gamma_a = 0.05$ (SGD). For the searching time functions $S_i(x)$, we further let $\delta = 10$, $\Delta = 1000$, and $\omega = 0.2$ in the piecewise-linear form defined in (22) and follow the function form stated in Section 3.1 for the bi-nominal $S_i(x)$.

With constant $\{\theta, k\}$, we now postulate $k = k^b = 40000$ (parking area/km) and investigate two cases with both excess AV parking $(\theta = 0.5 > \theta^b)$ and insufficient AV parking $(\theta = 0.25 < \theta^b)$. With $\theta = 0.5$, there are $m_a = 25000$, $m_c = 20000$ (parking spaces/km) and with $\theta = 0.25$, there are $m_a = 12500$, $m_c = 30000$ (parking spaces/km). The parking spans and distributions at equilibrium and LLP optimum are then delineated in Figure 2 with two parking searching time functions. The optimal differentiated parking pricings are also depicted in Figure 3. At equilibrium with piecewise-linear $S_i(x)$, $\{p_c, p_a\} = \{3.21, 1.90\}$ when $\theta = 0.5$ and $\{p_c, p_a\} = \{2.43, 2.12\}$ when $\theta = 0.25$. And at equilibrium with bi-nominal $S_i(x)$, $\{p_c, p_a\} = \{2.98, 1.84\}$ when $\theta = 0.5$ and $\{p_c, p_a\} = \{2.11, 2.03\}$ when $\theta = 0.25$.

From Figure 2, it can be observed that the trend of parking distributions along $x$ are similar between the two $S_i(x)$. In addition, Lemma 3.2 and Lemma 4.2 are verified that with constant $m_a, m_c$, the numbers of parked vehicles decrease with the distance to CBD at both equilibrium and LLP optimum. Furthermore, the parking span at LLP optimum is found to be larger than that at spatial equilibrium in the two cases not only for AVs as indicated in Proposition 4.1, but also for HVs. When it comes to the optimal parking pricing shown in Figure 3, it is worth noting that the optimal pricing can be negative, which can also be interpreted as the parking subsidy. Based on Figure 3, while the optimal pricing for both AVs and HVs is decreasing with $x$, the AV parking is actually subsidized to achieve the optimal parking distribution. It can further be verified that with optimal pricing, the total parking cost excluding pricing expense has reduced by 13.1% at $\theta = 0.5$ and 8.9% at $\theta = 0.25$ (with bi-nominal $S_i(x)$).

We next examine the ULP with the bi-nominal searching time function. Let $x = 5$ (km), $v_a = 50$ (SGD/area) and the land rent function follows $L(x) = 200\left(1 - \frac{x}{x}\right)$ (SGD/area) such that $L = 500$ (SGD). Again, we let the initial $k = k^b = 40000$. With unchanged $\varepsilon = 0.4$ and $\theta^b = 0.35$, the parking budget can be calculated as $\overline{NP} = 2.35 \times 10^7$ (SGD). The contour plot of the reduction percentage of total parking cost from benchmark case with different parking planning design is then depicted in Figure 5, where the parking budget constraint stated in (23) is highlighted in dashed. The indifference curve presents the parking planning designs with the same total parking cost as the benchmark case. Any design below such curve leads to higher total parking cost of travellers. In addition, the parking planning designs at benchmark, first- and second-best optimum are also highlighted, respectively. It can be found that the first- and second-best optimum are $\{\theta^{o1}, k^{o1}\} = \{0.15, 43744\}$ and $\{\theta^{o2}, k^{o2}\} = \{0.17, 20000\}$, which are in line with the theoretical results stated in Proposition 5.1 and Proposition 5.2. At $\varepsilon = 0.4$, the first- and second-best optimum can result in 9.2% and 5.7% reduction in total parking cost. The performance improvement from second- to first-best optimum is 3.6%. Furthermore, the maximum total parking cost reduction percentage from benchmark to first-best optimum is $\max \dfrac{TC^b - TC^{o1}}{TC^b} = 10.7\% < \left( \dfrac{\sqrt{v_a x + L}\sqrt{\phi} - \sqrt{L}\sqrt{\mu}}{\sqrt{v_a x + L}\sqrt{\phi} + \sqrt{L}\sqrt{\mu}} \right)^2 = 26.2\%$ at $\varepsilon = 0.61$, and that to second-best optimum is $\max \dfrac{TC^b - TC^{o2}}{TC^b} = 7.3\% < \left( \dfrac{\sqrt{\phi} - \sqrt{\mu}}{\sqrt{\phi} + \sqrt{\mu}} \right)^2 = 18.8\%$ at $\varepsilon = 0.66$. Lemma 5.2 and Lemma 5.3 are verified. Though the cost reduction is modest for the reason of already low $TC^b$ at benchmark case, as shown in Figure 5**Error! Reference source not found.**, inappropriate parking planning design can increase the total parking cost tremendously (with negative reduction percentage). In addition, the maximum performance improvement from second- to first-best optimum is $\max \dfrac{TC^{o2} - TC^{o1}}{TC^{o2}} = 3.8\% < \dfrac{v_a x \theta^b}{L + v_a x \theta^b} = 14.8\%$ at $\varepsilon = 0.51$, where Lemma 5.4 is verified. There is limited difference in system performance between the first- and second-best optimums. Instead, as shown in Figure 6, the second-best optimum can further save the infrastructure cost on parking by 7.4% at $\varepsilon = 0.4$.

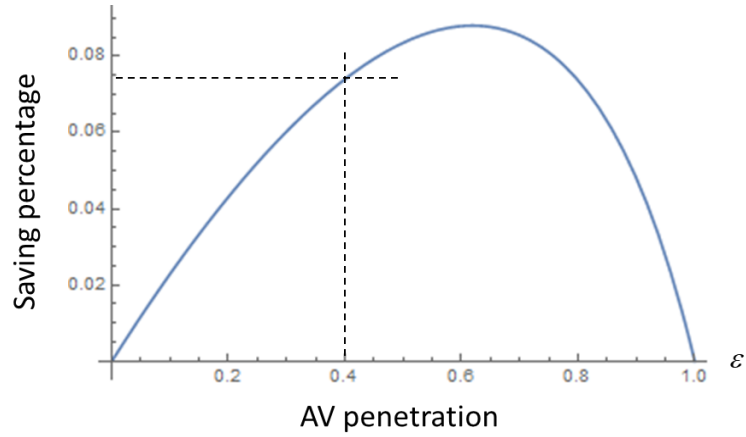

Figure 6 The saving percentage on parking budget at second-best optimum compared to benchmark case, at different penetration of AVs. (Numerical example)

Last but not least, we exemplify the general case with location-dependent parking supply. While $k(x) = 40000$ remains constant along the city with other parameters unchanged, we postulate $\theta(x) = \dfrac{\varepsilon}{1+e^{-4\left(x-\frac{1}{6}x\right)}}$ following a sigmoid function with $\theta(x) < \varepsilon = 0.4$ and $\theta\left(\dfrac{1}{6}x\right) = 0.25$, such that the parking areas close to city centre are mainly allocated for HV parking. For locations further away, the proportion of AV parking area increases yet is still less than AV penetration. Hence, the parking supply $m_i(x)$ $(i = a, c)$ varies along the city. The corresponding parking choice distributions of both AVs and HVs at equilibrium and optimum are then depicted in Figure 7. Compared to the case with constant $\theta = 0.25$ in Figure 2, while the number of parked HVs decreases with $x$ at both equilibrium and LLP optimum, it no longer holds for AVs. Due to the little AV parking supply near city centre, the number of parked AVs first increases then decreases with $x$. Though without constant $m(x)$, the parking spans at optimum are still found to be larger than those at spatial equilibrium for both AVs and HVs. Besides, it can be verified that $TC$ has decreased from 52611 to 49264 (SGD), whereas *NP* increases from $2.27 \times 10^7$ to $2.35 \times 10^7$ (SGD), with this location-dependent parking supply.

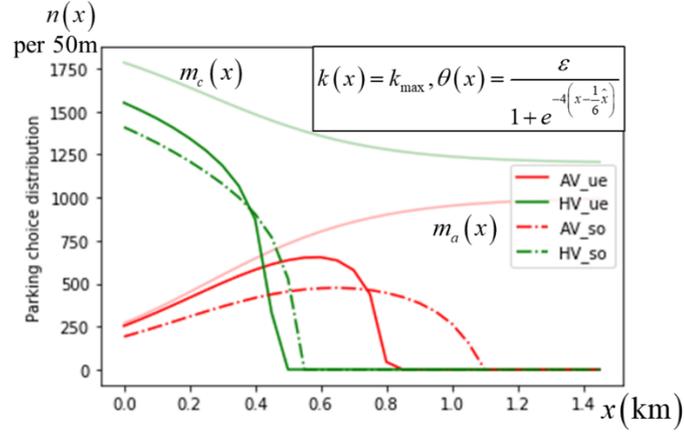

Figure 7 The parking distribution at equilibrium and optimum with given location-dependent parking supply (Numerical example).

### 7. **Concluding remarks**

In this paper, for the first time in literature the parking problem with mixed AVs and HVs is explicitly explored. The spatial equilibrium of parking location choices of both types of vehicles is investigated in a linear city corridor, with identifications of the different parking processes. Each type of parking is modelled with three components: HV driver's walking to city centre or AV's self-driving to parking location, the searching for parking at the determined parking location, and more particularly, the cruising for parking on the arterial road. Furthermore, with the aim to minimize the total parking cost of travellers in limited infrastructure expenditure, a bi-level PPDP is formulated to explore the optimal parking pricing and parking planning along the city. While optimal parking distribution is achieved with differentiated parking pricing at the LLP, total parking cost can be further reduced at optimal parking planning design for AVs and HVs.

Nevertheless, some extensions are expected to be further addressed. Foremost, we make the assumption of parking independence between AVs and HVs as there are different design and requirement of parking spaces for AVs and HVs (Nourinejad et al., 2018). However, it is possible that the parking spaces are shared by AVs and HVs. For instance, as the parking spaces for HVs are often in larger size with passages between rows, AVs may choose to park at the HVs' garage, particularly in the case with insufficient parking supply. It is worth studying the case with mixed usage of parking spaces in future studies. In that case, the mixed spatial parking location choices may be non-unique. Besides, though the differentiated parking pricing of AVs and HVs are proposed to optimize the parking distribution, it may raise equity issues and further research attention should be paid to promote the equity with the aid of, e.g., tradable credits (Wang et al., 2012). Last but not the least, it is interesting for future extensions to investigate PPDP not only in the continuum traffic corridor, but also in other discretized road networks.

Shladover, S. E., & Nowakowski, C. (2019). Regulatory challenges for road vehicle automation: Lessons from the California experience. *Transportation Research Part A: Policy and Practice, 122*, 125-133. doi:https://doi.org/10.1016/j.tra.2017.10.006

Simoni, M. D., Kockelman, K. M., Gurumurthy, K. M., & Bischoff, J. (2019). Congestion pricing in a world of self-driving vehicles: An analysis of different strategies in alternative future scenarios. *Transportation Research Part C: Emerging Technologies, 98*, 167-185. doi:https://doi.org/10.1016/j.trc.2018.11.002

Su, Q., & Wang, D. Z. W. (2019). Morning commute problem with supply management considering parking and ride-sourcing. *Transportation Research Part C: Emerging Technologies*. doi:https://doi.org/10.1016/j.trc.2018.12.015

Su, Q., & Wang, D. Z. W. (2020). On the morning commute problem with distant parking options in the era of autonomous vehicles. *Transportation Research Part C: Emerging Technologies, 120*, 102799. doi:https://doi.org/10.1016/j.trc.2020.102799

Szeto, W. Y., Jiang, Y., Wang, D. Z. W., & Sumalee, A. (2015). A Sustainable Road Network Design Problem with Land Use Transportation Interaction over Time. *Networks and Spatial Economics, 15*(3), 791-822. doi:10.1007/s11067-013-9191-9

Taxonomy and Definitions for Terms Related to Driving Automation Systems for On-Road Motor Vehicles. (2018). SAE International.

Tian, L.-J., Sheu, J.-B., & Huang, H.-J. (2019). The morning commute problem with endogenous shared autonomous vehicle penetration and parking space constraint. *Transportation Research Part B: Methodological, 123*, 258-278. doi:https://doi.org/10.1016/j.trb.2019.04.001

Tscharaktschiew, S., & Evangelinos, C. (2019). Pigouvian road congestion pricing under autonomous driving mode choice. *Transportation Research Part C: Emerging Technologies, 101*, 79-95. doi:https://doi.org/10.1016/j.trc.2019.02.004

van den Berg, V. A. C., & Verhoef, E. T. (2016). Autonomous cars and dynamic bottleneck congestion: The effects on capacity, value of time and preference heterogeneity. *Transportation Research Part B: Methodological, 94*, 43-60. doi:https://doi.org/10.1016/j.trb.2016.08.018

Wang, D. Z. W., & Lo, H. K. (2010). Global optimum of the linearized network design problem with equilibrium flows. *Transportation Research Part B: Methodological, 44*(4), 482-492. doi:10.1016/j.trb.2009.10.003

Wang, X., Yang, H., Zhu, D., & Li, C. (2012). Tradable travel credits for congestion management with heterogeneous users. *Transportation Research Part E:*
47

   *Logistics and Transportation Review, 48*(2), 426-437. doi:https://doi.org/10.1016/j.tre.2011.10.007

Yang, H., & H. Bell, M. G. (1998). Models and algorithms for road network design: a review and some new developments. *Transport Reviews, 18*(3), 257-278. doi:10.1080/01441649808717016

Yu, C., Sun, W., Liu, H. X., & Yang, X. (2019). Managing connected and automated vehicles at isolated intersections: From reservation- to optimization-based methods. *Transportation Research Part B: Methodological, 122*, 416-435. doi:https://doi.org/10.1016/j.trb.2019.03.002

Zhang, W., & Guhathakurta, S. (2017). Parking Spaces in the Age of Shared Autonomous Vehicles: How Much Parking Will We Need and Where? *Transportation Research Record, 2651*(1), 80-91. doi:10.3141/2651-09

Zhang, X., Liu, W., & Waller, S. T. (2019a). A network traffic assignment model for autonomous vehicles with parking choices. *Computer-Aided Civil and Infrastructure Engineering, 0*(0). doi:10.1111/mice.12486

Zhang, X., Liu, W., Waller, S. T., & Yin, Y. (2019b). Modelling and managing the integrated morning-evening commuting and parking patterns under the fully autonomous vehicle environment. *Transportation Research Part B: Methodological, 128*, 380-407. doi:https://doi.org/10.1016/j.trb.2019.08.010

**Appendix A. List of notation.**

*Parameters*

| | |
|---|---|
| $N$ | The total number of parkers. |
| $\varepsilon$ | The penetration rate of AVs, where the number of AV: $N_a = \varepsilon N$ and HV: $N_c = (1-\varepsilon) N$. |
| $\lambda_c$ | Walking cost per unit distance. |
| $\lambda_a$ | Aggregate self-driving cost per unit distance of AVs. |
| $\gamma_i$ | The unit time cost of parking searching of vehicle type $i(=a,c)$. |
| $\beta_i$ | The extra delay cost per cruising vehicle in type $i(=a,c)$. |
| $\phi$ | Relative size of unit AV parking space, with unit HV size equal to 1. |
| $v_a$ | Upgrade cost per unit area for AV parking |

| $x$ | Supply span of parking spaces |
|---|---|
| $k^b$ | Initial (benchmark) parking area per unit distance |
| $\overline{NP}$ | The monetary budget on parking infrastructure |

*Variables*

| $k(x)$ | Total area of parking spaces at location $x$. |
|---|---|
| $\theta(x)$ | The proportion of parking areas allocated to AVs at $x$. |
| $m_i(x)$ | Number of parking spaces for vehicle type $i(=a,c)$ at $x$, where $m_a(x) = \theta(x)k(x)/\phi$ and $m_c(x) = (1-\theta(x))k(x)$. |
| $\overline{x_i^j}$ | The parking span of vehicle type $i(=a,c)$, at case $j(=e,o)$ where $e$ indicates the unpriced spatial equilibrium, and $o$ indicates the social optimum. |
| $S_i(x)$ | Average searching time for parking of vehicle type $i(=a,c)$ at $x$. |
| $n_i^j(x)$ | The number of travellers choosing to park at location $x$ for vehicle type $i(=a,c)$, at case $j(=e,o)$. |
| $c_i(x)$ | The expected cruising cost for a parker in type $i(=a,c)$ at $x$. |
| $P_i(x)$ | The generalized parking cost of a parker in type $i(=a,c)$ at $x$. |
| $p_i$ | The equilibrium parking cost in type $i(=a,c)$. |
| $TP_i$ | The total parking cost of type $i(=a,c)$, excluding cruising. |
| $MP_i(x)$ | The marginal cost of parking in type $i(=a,c)$ at $x$. |
| $\tau_i(x)$ | The parking price of vehicle type $i(=a,c)$ at $x$. |
| $TCr$ | Total cruising cost of travellers |
| $TP^{\min}$ | The minimum total parking cost of travellers |
| TC | Total parking cost including parking pricing at LLP optimum |
| NP | Total aggregate cost on parking infrastructure |
| $L(x)$ | The unit area land rent of parking at $x$, and $L = \int_0^x L(u)du$. |

**Appendix B.   Proof of Lemma 5.2, Lemma 5.3 and Lemma 5.4**

We first prove Lemma 5.2. Foremost, the reduction percentage of total parking cost can be represented as

$$\frac{TC^b - TC^{o1}}{TC^b} = \frac{(1-\varepsilon)\varepsilon\left(\sqrt{\frac{v_a x + L}{L}}\phi - \sqrt{\mu}\right)^2 L}{(1-\varepsilon+\varepsilon\mu)\left(1-\varepsilon+\varepsilon\frac{v_a x + L}{L}\phi\right)L + \frac{(1-\omega)C(\varepsilon)\overline{NP}}{\lambda_c N^2}}.$$

It satisfies

$$\frac{TC^b - TC^{o1}}{TC^b} < \frac{(1-\varepsilon)\varepsilon\left(\sqrt{\frac{v_a x + L}{L}}\phi - \sqrt{\mu}\right)^2 L}{(1-\varepsilon+\varepsilon\mu)\left(1-\varepsilon+\varepsilon\frac{v_a x + L}{L}\phi\right)L}$$

as $\frac{(1-\omega)C(\varepsilon)\overline{NP}}{\lambda_c N^2} > 0$. The right-hand side of this inequality reach maximum $\left(\frac{\sqrt{v_a x + L}\sqrt{\phi} - \sqrt{L}\sqrt{\mu}}{\sqrt{v_a x + L}\sqrt{\phi} + \sqrt{L}\sqrt{\mu}}\right)^2$ at

$$\varepsilon = \frac{1}{1 + \sqrt{\frac{v_a x + L}{L}\phi\mu}}.$$

This completes the proof of Lemma 5.2. The proof of Lemma 5.3 is similar to that of Lemma 5.2 and it is omitted here to save space. Note that for Lemma 5.3, the critical AV penetration to reach the right-hand side of inequality is

$$\varepsilon = \frac{1}{1+\sqrt{\phi\mu}}.$$

As for Lemma 5.4, we let $\Delta = \sqrt{(L+v_a x)L} - L \in (0, v_a x)$ and

$$C_1 = \frac{(1-\omega)k^b C(\varepsilon)}{\lambda_c N^2} > 0,$$

there is

$$\frac{TC^{o2}-TC^{o1}}{TC^{o2}} = \frac{(1-\varepsilon)\varepsilon\sqrt{\phi}}{C_1+\left(1-\varepsilon+\sqrt{\phi}\sqrt{\mu\varepsilon}\right)^2}\frac{\left(2L(1-\varepsilon)\sqrt{\mu}+(L+\Delta+L)(1-\varepsilon-\varepsilon\mu)\sqrt{\phi}+2(\Delta+L)\varepsilon\sqrt{\mu\phi}\right)\Delta}{L^2(1-\varepsilon)+(\Delta+L)^2\varepsilon\phi}$$

. Further, it can be derived that $\partial\left(\frac{TC^{o2}-TC^{o1}}{TC^{o2}}\right)/\partial\mu < 0$ and hence when $\mu$ reaches its minimum, the reduction percentage achieves the maximum. Namely, it satisfy $\frac{TC^{o2}-TC^{o1}}{TC^{o2}} \leq \frac{(1-\varepsilon)^2}{C_1+(1-\varepsilon)^2}\frac{v_a x\theta^b}{L+v_a x\theta^b}$, where the right hand side is derived at $\mu \to 0$ and the equality holds if and only if $v_a \to 0$. Obviously, for $v_a > 0$

$\frac{(1-\varepsilon)^2}{C_1+(1-\varepsilon)^2}\frac{v_a x\theta^b}{L+v_a x\theta^b} < \frac{v_a x\theta^b}{L+v_a x\theta^b}$. This completes the proof $\square$